\documentclass[usenatbib,onecolumn]{mnras}
\usepackage{amsmath,amssymb}
\usepackage{graphicx,paralist,comment}

\let\Re\relax
\DeclareMathOperator{\Re}{Re}

\title{Imaging with a gravitational lens: the geometric view}

\author[V. T. Toth]{
Viktor T. Toth$^1$\\
$^1$Ottawa, Ontario K1N 9H5, Canada}

\date{Accepted XXX. Received YYY; in original form ZZZ}

\pubyear{2023}

\begin{document}

\label{firstpage}
\pagerange{\pageref{firstpage}--\pageref{lastpage}}

\maketitle

\begin{abstract}
We investigate imaging point sources with a monopole gravitational lens, such as the Solar Gravitational Lens in the geometric optics limit. We compute the light amplification of the lens used in conjunction with a telescope featuring a circular aperture that is placed in the focal region of the lens, compared to the amount of light collected by the same telescope unaided by a gravitational lens. We recover an averaged point-spread function that is in robust agreement with a wave-theoretical description of the lens, and can be used in practical calculations or simulations.
\end{abstract}


\section{Introduction}

In previous studies \citep{SGL2017,SGL2019b,SGL2020a,SGL2020c} we investigated imaging using a monopole gravitational lens, specifically the Solar Gravitational Lens (SGL) offering a comprehensive wave-optical treatment.

A key result in those studies was a form of the point spread function (PSF) of the lens that was averaged over the circular aperture of an observing telescope. It was with the help of this averaged PSF that we were able to calculate light amplification by the lens and also model both the images projected by the lens and also the images seen by the observing telescope.

In the present study, we re-derive the same results from first principles using only geometric optics. Apart from validating our previous results we hope that this treatment also makes the results more accessible. We must, of course, remind the reader that the geometric optics limit applies only when the wavelength is sufficiently short. In the case of the SGL, this implies that the quantity $kd\sqrt{2r_g/\bar{z}}\gg 1$, where $k$ is the wavenumber, $d$ is the observing telescope's aperture, $r_g$ is the Schwarzschild-radius of the lens and $\bar{z}$ is the distance from the lens to the observing telescope. The results presented here are valid, in particular, for observations made in optical or near-IR wavelengths using the SGL in combination with a meter-class optical telescope.

We define light amplification in this analysis by comparing the solid angle subtended by a telescope aperture that is unassisted by a gravitational lens against the solid angle through which a gravitational lens collects light that ultimately reaches the same telescope aperture. This is similar in principle to the frequently used concept of magnification in the gravitational lensing literature, comparing the solid angle subtended by a source when viewed directly vs. through a gravitational lens; but it is distinct in principle from the definition of light amplification used in wave optics, related to the magnitude of the Poynting-vector. Our goal, in part, was to demonstrate that the geometric analysis yields results that are compatible with the wave optical analysis.

This paper is organized as follows.
In Section~\ref{sec:monopole} we describe a standalone thin lens and compare it against the on-axis gravitational lens.
In Section~\ref{sec:offaxis} we move the thin lens aperture off-axis and derive our main result.
We present a brief discussion, relating this result to our earlier work, in Section~\ref{sec:disc}.

\section{The monopole gravitational lens}
\label{sec:monopole}

Our goal is to assess the amount of light from a distant point source intercepted by a circular telescope in the focal region of a gravitational lens. As a secondary objective, we also deduce the shape of the Einstein ring or Einstein arcs that are seen by this telescope as it looks at the lens. Given the geometry of the system, we rely on the paraxial approximation, characterized by distances $z$ much greater than the impact parameter $b$, $z\gg b$, thus for all angles $\theta={\cal O}(b/z)$, $\sin\theta\simeq \theta$.

The expression, ``light amplification'', is perhaps a misnomer. A true amplifier boosts a signal by injecting and modulating power from an external source. This is what amplifiers do in electronics: energy from a source (a battery pack, an external power supply) is modulated by the signal, faithfully replicating its behavior but at greater power levels.

This is not how a telescope (gravitational or otherwise) works. There is no external power source, no modulated signal. Rather, a telescope boosts signal levels simply by collecting more light. ``Light amplification'' thus boils down to a simple ratio of light collecting areas: the larger a telescope's aperture, the more light it collects.

Keeping this in mind, the question of light amplification thus becomes a relatively straightforward exercise in geometry: comparing the light collecting areas of different configurations.

Specifically, we wish to compare the light collecting area of an ordinary telescope with that of the same telescope but looking through (and boosted by) a gravitational lens. Therefore, we begin with describing an ordinary thin lens telescope with no gravitational lensing present.

\subsection{The thin lens telescope}
\label{sec:thin}

\begin{figure*}
\includegraphics{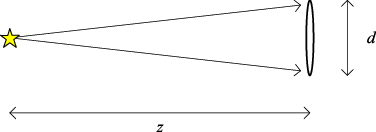}
\caption{\label{fig:thinlens}A thin lens telescope.}
\end{figure*}

First, let us consider a simple thin lens system with aperture $d$, at distance $z$ from a compact (point) source of light at infinity (Fig.~\ref{fig:thinlens}). This lens is characterized by the light collecting area $A_{\tt thin}=\pi(\tfrac{1}{2}d)^2$, at distance $z$ from the source. The solid angle that this aperture subtends is given by
\begin{align}
\Theta_{\tt thin}=\frac{A_{\tt thin}}{z^2}=\frac{\pi d^2}{4z^2}.
\label{eq:thinlens}
\end{align}

\subsection{Gravitational lens on-axis}
\label{sec:onaxis}

\begin{figure*}
\includegraphics{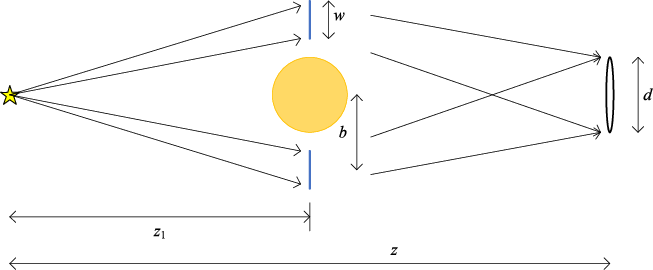}
\caption{\label{fig:lens}A gravitational lens.}
\end{figure*}

Let us compare this baseline case against that of a single gravitational lens, as depicted in Fig.~\ref{fig:lens}. The light collecting area of this lens is characterized by the annulus of the Einstein ring, with width $w$, radius $b$ (the impact parameter) at distance $z_1$ from the source.

\begin{figure*}
\includegraphics{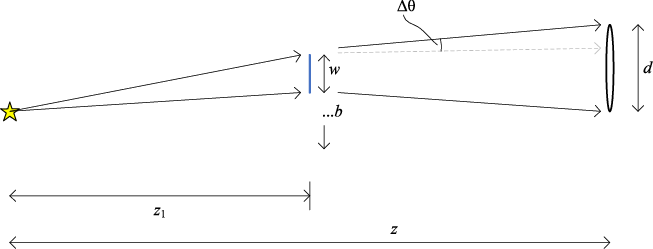}
\caption{\label{fig:divergence}Radial divergence of light rays.}
\end{figure*}

What is the value of $w$? It is tempting to think that it is simply the aperture $d$ scaled by the factor $z_1/z$, but that fails to take into account the slight radial divergence between light rays with impact parameter $b$ vs. $b'=b+w$. The well-known expression for the angle characterizing the gravitational deflection of light is $\theta=2r_g/b$ \citep{Einstein1915}, where $r_g$ is the Schwarzschild-radius of the lens. If we associate this impact parameter with the innermost light ray that hits the observing aperture, the outermost light ray will, in turn, be deflected by\footnote{Throughout this analysis, we assume that all angles are very small, therefore $\sin\theta\simeq\theta$ is always a permissible approximation.} (see also Fig.~\ref{fig:divergence}):
\begin{align}
\theta'=\frac{2r_g}{b'}=\frac{2r_g}{b}\frac{b}{b+w}\simeq \frac{2r_g}{b}\left(1-\frac{w}{b}\right)=\theta\left(1-\frac{w}{b}\right),
\label{eq:theta'}
\end{align}
thus $\Delta\theta=\theta w/b$.

The relationship between $d$ and $w$ can be readily read off Fig.~\ref{fig:divergence}:
\begin{align}
d=w\frac{z}{z_1}+(z-z_1)\Delta\theta=w\left(\frac{z}{z_1}+(z-z_1)\frac{\theta}{b}\right).
\end{align}

Thus, given a lens aperture $d$, the corresponding width of the Einstein ring is given by $w$, and the light collecting area of the Einstein ring annulus can be calculated as
\begin{align}
A_{\tt ER}=2\pi bw=2\pi bd\left(\frac{z}{z_1}+(z-z_1)\frac{\theta}{b}\right)^{-1}.
\end{align}

The impact parameter itself can be determined by noting that a gravitational lens deflects light by the angle $\theta=2r_g/b$ where $r_g=2GM/c^2$ is the Schwarzschild radius of the lens. In this case, the lens must deflect a light ray diverging from the optical axis by the angle $\arcsin b/z_1\sim b/z_1$ into a ray that approaches the optical axis by the angle $\arcsin b/(z-z_1)\sim b/(z-z_1)$ (where we assumed that the impact parameter in all cases is much smaller than the distances along the optical axis), thus
\begin{align}
\left(\frac{1}{z_1}+\frac{1}{z-z_1}\right)b=\frac{2r_g}{b},
\end{align}
or
\begin{align}
\label{eq:b}
b=\sqrt{\frac{2r_gz_1(z-z_1)}{z}}.
\end{align}
Therefore,
\begin{align}
A_{\tt ER}=2\pi bd\left(\frac{z}{z_1}+(z-z_1)\frac{2r_g}{b^2}\right)^{-1}=2\pi bd\left(\frac{z}{z_1}+(z-z_1)\frac{2r_g z}{2r_g z_1(z-z_1)}\right)^{-1}
=\pi bd\frac{z_1}{z}.
\label{eq:Aer}
\end{align}
With respect to the source, from which it is located at the distance $z_1$, the Einstein ring subtends the solid angle
\begin{align}
\Theta_{\tt ER}=\frac{A_{\tt ER}}{z_1^2}=\frac{\pi bd}{zz_1}.
\end{align}
The light amplification of the gravitational lens, compared to the light collected by a thin lens telescope with aperture $d$ at the same distance $z$ from the source, is given by
\begin{align}
\mu_{\tt ER}=\frac{\Theta_{\tt ER}}{\Theta_{\tt thin}}=\frac{\pi bd}{zz_1} \frac{4z^2}{\pi d^2}=\frac{z}{z_1} \frac{4b}{d}.
\label{eq:muER}
\end{align}
In the limiting case when $z\simeq z_1$ (i.e., when the source distance is much greater than the distance to the focal region), this expression simplifies to $4b/d$ with $b=\sqrt{2r_g(z-z_1)}$. This result is in agreement with Eq.~(143) in \cite{SGL2017} in the asymptotic limit of large arguments ($\lambda\to 0$) for the Bessel functions therein.

\section{Thin lens off-axis}
\label{sec:offaxis}

In the preceding derivation, we assumed an axisymmetric system: the point source, the center of the gravitational lens and the optical axis of the observing telescope were lined up. We are now interested in the amount of light that the telescope collects when it is moved in the image plane away from the principal axis of the gravitational lens.

\begin{figure*}
\includegraphics[scale=0.5]{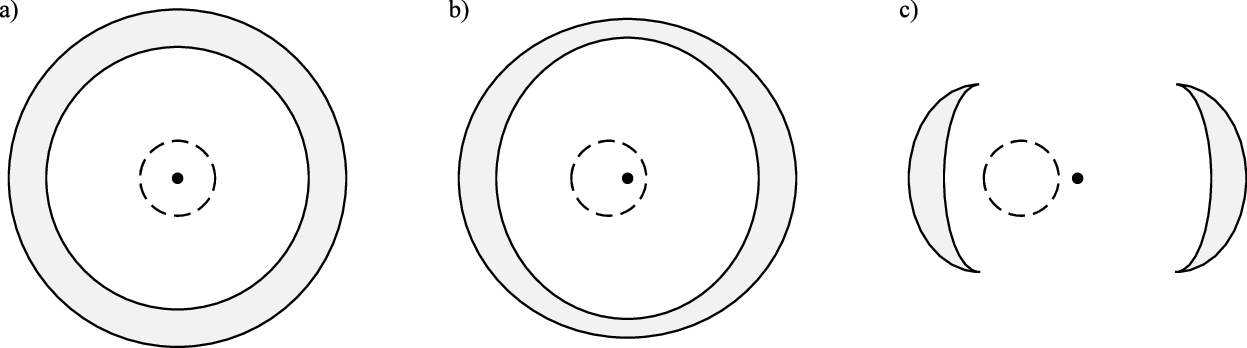}
\caption{\label{fig:chords}Axial view of moving the observing aperture (dashed line, not to scale) off-axis and the effect on the view of the Einstein ring (shaded area): a) aperture is exactly on the primary axis of the gravitational lens; b) telescope is slightly off-axis but the axis still falls within the aperture; c) telescope is outside the axis.}
\end{figure*}

\begin{figure*}
\includegraphics[scale=0.5]{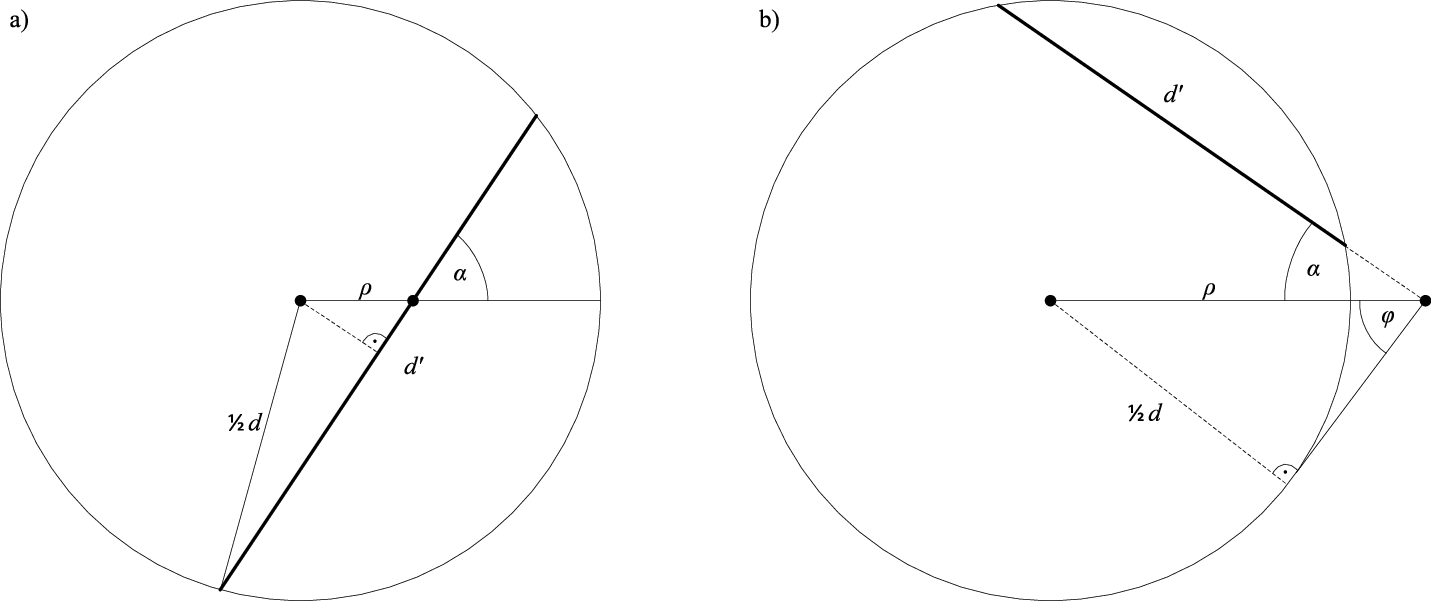}
\caption{\label{fig:inandout}The geometry of the off-axis aperture a) with the primary axis within the aperture, and b) with the primary axis outside the aperture.}
\end{figure*}

These cases are illustrated in Fig.~\ref{fig:chords}. As a result of moving the telescope off-axis, radial light rays are now separated not by the full diameter of the aperture but by the corresponding chord, the length of which depends on direction. The case when the aperture if off-axis but the primary axis of the gravitational lens is still within the aperture is further illustrated by Fig.~\ref{fig:inandout}a.

From this figure, we can read off the dependance of the chord length $d'$ on the angle $\alpha$:
\begin{align}
d'(\alpha)=2\sqrt{(\tfrac{1}{2}d)^2-\rho^2\cos^2\alpha}.
\end{align}
The width of the Einstein ring is
\begin{align}
w(\alpha)=\frac{1}{2}\frac{z_1}{z}d'(\alpha)=\frac{z_1}{z}\sqrt{(\tfrac{1}{2}d)^2-\rho^2\cos^2\alpha}.
\label{eq:w1}
\end{align}
To calculate the light collecting area of the Einstein ring, we need to compute the following integral:
\begin{align}
A_{\tt ER}=b\int_0^{2\pi}d\alpha w(\alpha)=b\frac{z_1}{z}\int_0^{2\pi}d\alpha \sqrt{(\tfrac{1}{2}d)^2-\rho^2\cos^2\alpha}=2bd\frac{z_1}{z}~{\tt E}\left(\frac{\rho}{\tfrac{1}{2}d}\right).
\end{align}
where ${\tt E}(z)$ is the complete elliptic integral of the second kind in the Legendre normal form.

In the limit $\rho\to 0$, we obtain the limit $A_{\tt ER}=\pi bdz_1/z$, exactly in accordance with (\ref{eq:Aer}).

Moving the aperture further off-axis, we reach the region where the primary axis of the gravitational lens no longer falls within the aperture. This case is illustrated in Fig.~\ref{fig:inandout}b.

In this case, the aperture no longer sees a full Einstein ring. Light appears only from the directions that fall within the range of angles $\pm\phi$ and $\pi\pm\phi$, as indicated in the figure. The width of the Einstein ring is given by the corresponding chords:
\begin{align}
w(\alpha)=\frac{1}{2}\frac{z_1}{z}d'(\alpha)=\frac{z_1}{z}\sqrt{(\tfrac{1}{2}d)^2-\rho^2\sin^2\alpha},
\label{eq:w2}
\end{align}
where ${\tt E}(z,k)$ is the incomplete elliptic integral of the second kind in Legendre normal form. Correspondingly, we have
\begin{align}
A_{\tt ER}=4b\int_0^\phi d\alpha w(\alpha)=2bd\frac{z_1}{z}~{\tt E}\left(\frac{\tfrac{1}{2}d}{\rho},\frac{\rho}{\tfrac{1}{2}d}\right)
\end{align}

Noting a known relationship between the complete and incomplete elliptic integrals of the second kind for positive real arguments, namely that
\begin{align}
\Re{\tt E}(x)=\Re{\tt E}\left(\frac{1}{x},x\right),
\end{align}
which is to say that, for $x\ge 0$, $x\in \mathbb{R}$,
\begin{align}
\Re\int_0^1d\alpha\frac{\sqrt{1-\alpha^2x^2}}{\sqrt{1-\alpha^2}}=\Re\int_0^{1/x}d\alpha\frac{\sqrt{1-\alpha^2x^2}}{\sqrt{1-\alpha^2}},
\end{align}
we can write
\begin{align}
A_{\tt ER}=2bd\frac{z_1}{z}~\Re{\tt E}\left(\frac{\rho}{\tfrac{1}{2}d}\right),
\end{align}
which is valid for all values of $\rho$.

As before, the corresponding solid angle is calculated as
\begin{align}
\Theta_{\tt ER}=\frac{A_{\tt ER}}{z_1^2}=\frac{2bd}{zz_1}~\Re{\tt E}\left(\frac{\rho}{\tfrac{1}{2}d}\right),
\end{align}

Finally, the corresponding light amplification factor can be determined by comparing against a circular aperture with diameter $d$ that is not assisted by a gravitational lens, as we did in (\ref{eq:muER}):
\begin{align}
\mu_{\tt ER}=\frac{\Theta_{\tt ER}}{\Theta_{\tt thin}}=
\frac{A_{\tt ER}}{z_1^2}=\frac{2bd}{zz_1}~\Re{\tt E}\left(\frac{\rho}{\tfrac{1}{2}d}\right)
\frac{4z^2}{\pi d^2}=\frac{8bz}{\pi z_1 d}~\Re{\tt E}\left(\frac{\rho}{\tfrac{1}{2}d}\right)
\end{align}
that, in the limit $\rho\to 0$, reduces to (\ref{eq:muER}).

\begin{figure*}
\includegraphics[scale=0.4]{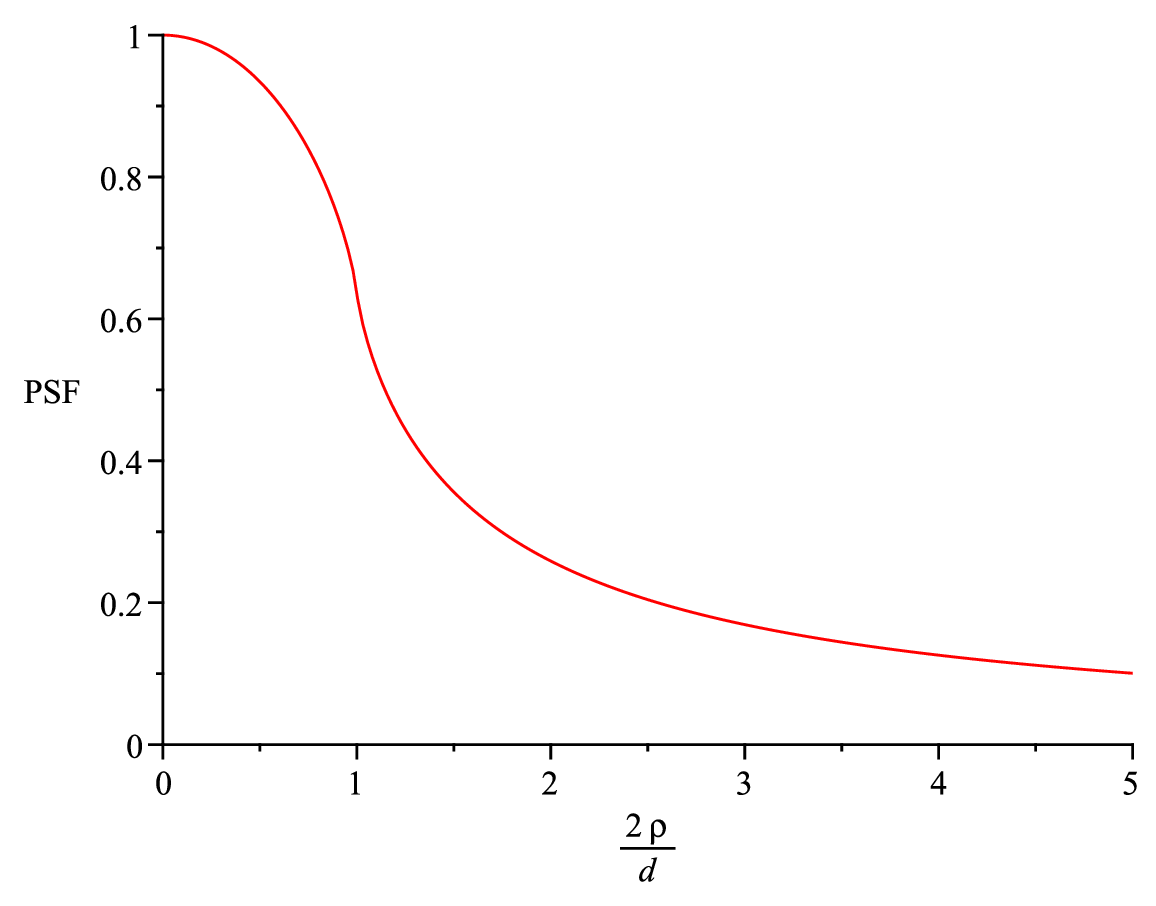}
\caption{\label{fig:PSF}The gravitational lens PSF as averaged over a circular aperture.}
\end{figure*}

Correspondingly, the normalized ``averaged PSF'' of the gravitational lens, with respect to a circular observing aperture with diameter $d$, is given by
\begin{align}
{\tt PSF}(\rho)=\frac{2}{\pi}~\Re{\tt E}\left(\frac{\rho}{\tfrac{1}{2}d}\right),
\label{eq:PSF}
\end{align}
also shown in Fig.~\ref{fig:PSF}, and in precise agreement with results we first obtained in \cite{SGL2020a,SGL2020c}.

\section{Discussion}
\label{sec:disc}

In this paper, relying entirely on geometric optics, we recovered two fundamental results concerning gravitational lenses.

First, we established that the light amplification of a monopole gravitational lens, used in conjunction with an observing telescope, in comparison with using the same observing telescope for direct observation, is given by $4b/d$ where $b$ is the gravitational lens impact parameter and $d$ is the telescope aperture. This result is in robust agreement with the result that we derived in earlier work using a theoretically robust wave-optical description of the relativistic lens.

Second, we established that the wavelength-independent ``averaged'' PSF of the lens, applicable in the case when the observing telescope aperture is much larger than the spatial wavelength of the diffraction pattern of the lens in the image plane, is given, in terms of the incomplete elliptic integral of the second kind in Legendre normal form, by ${\tt PSF}(\rho)=\tfrac{2}{\pi}~\Re{\tt E}(d/2\rho,2\rho/d)$. Again, this result is in perfect agreement with the derivation that we accomplished in earlier work using a wave-theoretical description of the lens.

In our analysis, we treated the gravitational lens as a monopole. Actual physical gravitational lenses are not perfect monopoles \citep{SGL2021a}. While it is possible in principle to incorporate the effect of the quadrupole moment (which effectively displaces the optical axis by an amount proportional to $\sin 2\alpha$, resulting in a modulation of the width of the Einstein ring or breaking it up into an Einstein cross depending on the quadrupole moment's magnitude) it is unlikely that such a description would lead to an integrable expression as in the monopole case. We have not attempted this derivation.

The result (\ref{eq:PSF}) is important. It accurately describes gravitational lenses in the limit of geometric optics in numerous scenarios, including many microlensing cases, where higher multipole moments can be neglected. We hope that our derivation of this result from first principles using geometric optics increases confidence in its validity.

\section*{Acknowledgments}

VTT thanks Slava Turyshev for discussions and acknowledges the generous support of  David H. Silver, Plamen Vasilev and other Patreon patrons.

\section*{Data Availability}

No data was generated and/or analysed to produce this article.

\bibliographystyle{mnras}
\bibliography{SGLGEOM}

\begin{thebibliography}{}
\makeatletter
\relax
\def\mn@urlcharsother{\let\do\@makeother \do\$\do\&\do\#\do\^\do\_\do\%\do\~}
\def\mn@doi{\begingroup\mn@urlcharsother \@ifnextchar [ {\mn@doi@}
  {\mn@doi@[]}}
\def\mn@doi@[#1]#2{\def\@tempa{#1}\ifx\@tempa\@empty \href
  {http://dx.doi.org/#2} {doi:#2}\else \href {http://dx.doi.org/#2} {#1}\fi
  \endgroup}
\def\mn@eprint#1#2{\mn@eprint@#1:#2::\@nil}
\def\mn@eprint@arXiv#1{\href {http://arxiv.org/abs/#1} {{\tt arXiv:#1}}}
\def\mn@eprint@dblp#1{\href {http://dblp.uni-trier.de/rec/bibtex/#1.xml}
  {dblp:#1}}
\def\mn@eprint@#1:#2:#3:#4\@nil{\def\@tempa {#1}\def\@tempb {#2}\def\@tempc
  {#3}\ifx \@tempc \@empty \let \@tempc \@tempb \let \@tempb \@tempa \fi \ifx
  \@tempb \@empty \def\@tempb {arXiv}\fi \@ifundefined
  {mn@eprint@\@tempb}{\@tempb:\@tempc}{\expandafter \expandafter \csname
  mn@eprint@\@tempb\endcsname \expandafter{\@tempc}}}

\bibitem[\protect\citeauthoryear{{Einstein}}{{Einstein}}{1915}]{Einstein1915}
{Einstein} A.,  1915, Sitzungsberichte der K{\"o}niglich Preussischen Akademie
  der Wissenschaften, \href
  {https://ui.adsabs.harvard.edu/abs/1915SPAW.......831E} {pp 831--839}

\bibitem[\protect\citeauthoryear{Turyshev \& Toth}{Turyshev \&
  Toth}{2017}]{SGL2017}
Turyshev S.~G.,  Toth V.~T.,  2017, \mn@doi [Phys. Rev. D]
  {10.1103/PhysRevD.96.024008}, 96, 024008

\bibitem[\protect\citeauthoryear{Turyshev \& Toth}{Turyshev \&
  Toth}{2019}]{SGL2019b}
Turyshev S.~G.,  Toth V.~T.,  2019, \mn@doi [Phys. Rev. D]
  {10.1103/PhysRevD.100.084018}, 100, 084018

\bibitem[\protect\citeauthoryear{Turyshev \& Toth}{Turyshev \&
  Toth}{2020a}]{SGL2020a}
Turyshev S.~G.,  Toth V.~T.,  2020a, \mn@doi [Phys. Rev. D]
  {10.1103/PhysRevD.101.044025}, {101}, {044025}

\bibitem[\protect\citeauthoryear{Turyshev \& Toth}{Turyshev \&
  Toth}{2020b}]{SGL2020c}
Turyshev S.~G.,  Toth V.~T.,  2020b, \mn@doi [Phys. Rev. D]
  {10.1103/PhysRevD.102.024038}, {102}, {024038}

\bibitem[\protect\citeauthoryear{Turyshev \& Toth}{Turyshev \&
  Toth}{2021}]{SGL2021a}
Turyshev S.~G.,  Toth V.~T.,  2021, \mn@doi [Phys. Rev. D]
  {10.1103/PhysRevD.103.064076}, 103, 064076

\makeatother
\end{thebibliography}

\label{lastpage}

\end{document}